\documentclass[10pt,A4paper,amsmath,amssymb,aps,etoolbox,floatfix,longbibliography,nofootinbib,
showkeys,noshowpacs,superscriptaddress,twocolumn]{revtex4}

\newcommand{\bea}{\begin{eqnarray}}
\newcommand{\beq}{\begin{equation}}

\newcommand{\eea}{\end{eqnarray}}
\newcommand{\eeq}{\end{equation}}
\newcommand{\ds}{\displaystyle}

\usepackage{amsmath}
\usepackage{amsfonts}
\usepackage{amssymb}
\usepackage{bm}       
\usepackage[usenames,dvipsnames]{color}
\usepackage{comment}
\usepackage{dcolumn}  
\usepackage{epsfig}
\usepackage{graphics}
\usepackage{graphicx} 
\usepackage{lipsum}
\usepackage{natbib}
\usepackage{ragged2e}
\usepackage[rightcaption]{sidecap}

\usepackage{placeins} 
\usepackage{psfrag}
\usepackage{times}
\usepackage[varg]{txfonts}
\usepackage{xcolor}
\usepackage{xspace}

\graphicspath{%
    {converted_graphics/}
    {/}
}

\begin{document}

\title{Photon frequency variation in non-linear electro-magnetism}

\author{Alessandro D.A.M. Spallicci \footnote{email: spallicci@cnrs-orleans.fr 
}}

\affiliation{\mbox{Laboratoire de Physique et Chimie de l'Environnement et de l'Espace (LPC2E) UMR 7328}\\
\mbox{Centre National de la Recherche Scientifique (CNRS), Universit\'e d'Orl\'eans (UO), Centre National d'\'Etudes Spatiales (CNES)}\\
\mbox {3A Avenue de la Recherche Scientifique, 45071 Orl\'eans, France}}

\affiliation{\mbox{Observatoire des Sciences de l'Univers en region Centre (OSUC) UMS 3116} \\
\mbox{ Universit\'e d'Orl\'eans (UO), Centre National de la Recherche Scientifique (CNRS)}
\mbox{ Observatoire de Paris (OP), Universit\'e Paris Sciences \& Lettres (PSL)}\\
\mbox{1A rue de la F\'{e}rollerie, 45071 Orl\'{e}ans, France}}

\affiliation{\mbox{UFR Sciences et Techniques, 
Universit\'e d'Orl\'eans, Rue de Chartres, 45100 Orl\'{e}ans, France}}

\author{Abedennour Dib \footnote{email: abedennour.dib@cnrs-orleans.fr }}

\affiliation{\mbox{Laboratoire de Physique et Chimie de l'Environnement et de l'Espace (LPC2E) UMR 7328}\\
\mbox{Centre National de la Recherche Scientifique (CNRS), Universit\'e d'Orl\'eans (UO), Centre National d'\'Etudes Spatiales (CNES)}\\
\mbox {3A Avenue de la Recherche Scientifique, 45071 Orl\'eans, France}}

\affiliation{\mbox{Observatoire des Sciences de l'Univers en region Centre (OSUC) UMS 3116} \\
\mbox{ Universit\'e d'Orl\'eans (UO), Centre National de la Recherche Scientifique (CNRS)}
\mbox{ Observatoire de Paris (OP), Universit\'e Paris Sciences \& Lettres (PSL)}\\
\mbox{1A rue de la F\'{e}rollerie, 45071 Orl\'{e}ans, France}}

\affiliation{\mbox{UFR Sciences et Techniques, 
Universit\'e d'Orl\'eans, Rue de Chartres, 45100 Orl\'{e}ans, France}}

\author{Jos\'e A. Helay\"el-Neto \footnote{email: helayel@cbpf.br}}

\affiliation{\mbox{Departamento de Astrof\'{\i}sica, Cosmologia e Intera\c{c}\~{o}es Fundamentais (COSMO), Centro Brasileiro de Pesquisas F\'{\i}sicas (CBPF)}\\
\mbox {Rua Xavier Sigaud 150, 22290-180 Urca, Rio de Janeiro, RJ,  Brazil}}

\date{6 April 2024}

\begin{abstract}
We set a generalised non-linear Lagrangian, encompassing Born-Infeld and Heisenberg-Euler theories among others. The Lagrangian reduces to the Maxwell Lagrangian at lowest order. The field is composed by a propagating light-wave in an electro-magnetic background. The wave exhibits energy variation when the background is space-time dependent. In the photon description, this implies a red or a blue shift, like what we obtained in massive theories, as the de Broglie-Proca or effective mass theories as the Standard-Model Extension under Lorentz symmetry violation. The two results, photon energy-conservation and the frequency shift are instead new for non-linear electro-magnetism. We conclude by discussing how these static frequency {shifts} when added to the expansion red shift allow new interpretations in cosmology or for atomic spectra. We finally consider the consequences on the Poincar\'e symmetry. 
\end{abstract}


\maketitle

\section{Introduction.} 

Non-Linear Electro-Magnetism (NLEM) was proposed first by Born and Infeld (BI) 
\cite{born-infeld-1934a,born-infeld-1935} for regularising point charges and by Heisenberg and Euler (HE) \cite{heisenberg-euler-1936} for dealing with strong fields. Both theories are used for second order Quantum Electro-Dynamics (QED) \cite{fouchebattestirizzo2016}. While the HE model is nowadays thoroughly derivable through QED loop calculations, it is still a very useful tool to also study vacuum birefringence and it is actively used by various collaborations to analyse through strong magnetic field based experiments vacuum polarisation, constancy of the speed of light and the nature of vacuum \cite{Ejilli-etal-2020,Agil-Battesti-Rizzo-2022,Agil-Battesti-Rizzo-2023,Robertson-etal-2021,Kraych-etal-2024}. The QED predicted value of birefringence was only a factor seven away from the experimental confirmation \cite{Ejilli-etal-2020}. Meanwhile, the BI model found a renewed interest from the community as it naturally emerges as a low-energy limit of certain string theories \cite{Tseytlin-2000}.

For a review including more recent NLEM theories, see \cite{Sorokin-2022}. The non-linear effects of electro-magnetism are investigated experimentally also through interferometry \cite{Schellestede-Perlick-Lammerzahl-2015} and colliders for photon splitting \cite{Akhmadaliev-etal-2002}, photon-photon interactions \cite{denterria-dasilveira-2013,Pike-Mackenroth-Hill-Rose-2014,aaboud-et-al-2017}, analysed with respect to the BI theory \cite{Ellis-Mavromatos-You-2017,Ellis-Mavromatos-Roloff-You-2022}.  

In the Standard-Model (SM) Extension (SME) under the Lorentz Symmetry Violation (LSV), 
an effective mass emerges and its value is related to the LSV vector or tensor \cite{bonetti-dossantosfilho-helayelneto-spallicci-2017,bonetti-dossantosfilho-helayelneto-spallicci-2018}. There occurs also an energy variation of the light-wave propagating in an Electro-Magnetic (EM) and LSV backgrounds \cite{bonetti-dossantosfilho-helayelneto-spallicci-2018}. The variation of the light-wave energy contains a term that couples to the LSV background and to an EM constant background. 

In massive theories as de Broglie-Proca (dBP) \cite{debroglie-1923,debroglie-1936,proca-1936d,proca-1937,db40}, the variation of the energy contains a term which again does not require the EM background to be space-time dependent. Even if the field is constant, the associated potential is not, and this condition suffices. 
Energy varies also in Maxwell's theory if the background is space-time dependent \cite{spallicci-sarracino-capozziello-2022}.

Herein, we turn to Generalised NLEM (GNLEM) and work out the frequency shift, while in future work, we deal with the emergence of a mass in GNLEM. Photon masses are differently tested in the solar wind \cite{retino-spallicci-vaivads-2016} and through Fast Radio Bursts \cite{boelmasasgsp2016,boelmasasgsp2017}. Herein, we work in absence of LSV and of any pre-defined mass and ask under which conditions {\it in vacuo} the photon energy would not be conserved. 

As we did for the dBP and SME cases, we use the correspondence wave-particle, down to a single photon \cite{aspect-grangier-1987} and derive the energy-momentum tensor variation of the photon and therefore a frequency shift. We consider the EM background non-dynamical and we do not describe its equations of motion. We figure the photon as a small perturbation of the EM background. 
In future work, we will tackle the issue of a dynamical background, imposing that its energy-momentum tensor variation equates that of the photon. 

Finally, the frequency shift is towards the blue or the red. This $z_{\rm S}$ shift is static by nature and it is not associated to the universe expansion. But if added to the expansion red shift, it spins-off new interpretations in cosmology, thanks to the recasting of the observed $z$ shift. We applied the recasting for examining alternatives to the accelerated expansion. Indeed, by combining the static and the expansion shifts, the red shift and luminosity distances of Supernovae Ia (SNeIa) get in agreement, without evoking dark energy \cite{helayelneto-spallicci-2019,spallicci-etal-2021,spallicci-sarracino-capozziello-2022,sarracino-spallicci-capozziello-2022}. The findings of the James Webb Space telescope appear to be in tension with $\Lambda$CDM cosmology \cite{Gupta-2023} and again applying the recasting might alleviate or cancel the tension.

The GNLEM photon energy variation has a large range of orders of magnitude. The uncertainty is due to a very difficult assessment, requiring not only the knowledge of the magnetic fields (strength and direction) for each of the galaxies and of the intergalactic spaces crossed by the photon, but also knowledge of their respective alignments.               

\section{Generalised Non-Linear Electro-Magnetism} 

We build a general Lagrangian 
${\cal L} = {\cal L} ({\cal F},{\cal G})$ [SI units],
as polynomial, function of integer powers of the quadratic fields $({\cal F}$ and its dual ${\cal G})$. 

For the sake of generality, we consider the Lagrangian density to be built up in terms of both the invariants ${\cal F},{\cal G}$. The choice of using also the parity violating ${\cal G}$ term is adopted for the purpose of generality. It is nonetheless important to underline that the Lagrangian and the field equations we shall be working with, i.e. Eqs. (\ref{lagrangian}, \ref{feq2}), do not break the parity invariance of electrodynamics because we are dealing with a purely photonic model, and parity violations occurs in the Fermionic interaction sector. 
To consider parity preserving phenomena we could simply consider only even powers of G at the cost of generality.
 We also would like to point out that the ${\cal G}$-terms - and not only the ${\cal F}$-terms - may be induced by some fundamental physics, as it is case with the BI Lagrangian, which naturally appears in the low-energy (field-theoretical) limit of string theories  \cite{Tseytlin-2000}.

For $\vec {E}$ and $\vec {B}$, the electric and magnetic~field, respectively; $F^{\sigma\tau}$ and 
${\tilde {F}}^{\sigma\tau} = G^{\sigma\tau}$ the electro-magnetic field tensor and its dual, respectively; ${A}^\sigma$ the~4-potential for 
$\phi$ time and ${\vec {A}}$ space components; the vacuum permeability $\mu_0$ 
and the speed of light $c$
; $\epsilon_{\sigma\tau\kappa\lambda}$ the Levi Civita pseudo-tensor, we have

\begin{align}
& {\cal F} \!=\! {\ds -\frac{1}{4\mu_0}} {F}^2 \!=\! {\ds -\frac{1}{4\mu_0}}{F}_{\sigma\tau}{F}^{\sigma\tau} \!=\! {\ds \frac{1}{2\mu_0}}
\left({\ds \frac{{\vec { E}^2}}{c^2}} - \vec { B}^2\right)~,\label{f}\\
& {\cal G} \!=\! -{\ds \frac{1}{4\mu_0}} {F}_{\sigma\tau} {\tilde {F}}^{\sigma\tau} \!=\! -{\ds \frac{1}{4\mu_0}} {F}_{\sigma\tau} G^{\sigma\tau} \!=\! {\ds \frac{1}{\mu_0}}{\ds \frac {{\vec {E}}}{c}}\cdot{\vec {B}}~,\label{g}\\
& F_{\sigma\tau} = \partial_\sigma {A}_\tau - \partial_\tau {A}_\sigma~,~~~
F^{\sigma\tau} = \partial^\sigma {A}^\tau - \partial^\tau {A}^\sigma~,\\
& {\tilde {F}}_{\sigma\tau} \!=\! G_{\sigma\tau}\!=\!{\ds\frac{1}{2}} \epsilon_{\sigma\tau\kappa\lambda}F^{\kappa\lambda}~,~ 
{\tilde {F}}^{\sigma\tau} \!=\! G^{\sigma\tau}\!=\!{\ds\frac{1}{2}} \epsilon^{\sigma\tau\kappa\lambda}F_{\kappa\lambda}~,\\
& {A}^\sigma = \left({\ds \frac{\phi}{c}}, {\vec A}\right)~,~~~{A}_\sigma = \left({\ds\frac{\phi}{c}}, - {\vec {A}}\right)~.
\end{align} 

We set our scenario where the total $F$ ($G$) fields are composed by the background $F_{\rm B}$ ($G_{\rm B}$) and the photon $f$ ($g$) fields

\begin{align}
\!\!\!\!F^{\mu\nu} = F_{\rm B}^{\mu\nu} + f^{\mu\nu}, ~~~ G^{\mu\nu} = G_{\rm B}^{\mu\nu} + g^{\mu\nu} = {\tilde F}_{\rm B}^{\mu\nu} + 
{\tilde f}^{\mu\nu}~. 
\label{split}
\end{align}

The Minkowski metric $\eta$ signature is (+ - - -). The Greek (Latin) indices run over the four space-time (three space) dimensions. We drop momentarily $\mu_0$ for the sake of space to pick it up again for the main results. 

\begin{widetext}

\section{The photon energy-momentum tensor}

We rework Eqs. (\ref{f},\ref{g}) by applying the position (\ref{split}) 

\flushleft{
\begin{align}
&{\cal F}\!=\!
\underbrace{-\frac{1}{4} F^2_{\rm B}}_{{\cal F}_{\rm B}} - \frac{1}{2} F_{\rm B} f 
\underbrace{-\frac{1}{4} f^2}_{{\cal F}_{\rm f}}
\!=\! 
{\cal F}_{\rm B} \underbrace{- \frac{1}{2} F_{\rm B} f  + {\cal F}_{\rm f}}_{\delta {\cal F}}
\!=\!
{\cal F}_{\rm B} + \delta {\cal F},\\
&{\cal G}
\!=\! 
\underbrace{-\frac{1}{4} F_{\rm B} {\tilde F}_{\rm B}}_{{\cal G}_{\rm B}} 
- \frac{1}{2} {\tilde F}_{\rm B} f 
\underbrace {\!-\frac{1}{4} f {\tilde f}}_{{\cal G}_{\rm f}} 
\!=\! 
 {\cal G}_{\rm B} \underbrace {\!-\! \frac{1}{2} {\tilde F}_{\rm B} f \!+\! {\cal G}_{\rm f}}_{\delta {\cal G}} 
\!=\! 
{\cal G}_{\rm B}\!+\!\delta {\cal G}~,
\end{align}
}
having used the identity $F_{\rm B} {\tilde f} = {\tilde F}_{\rm B} f$.
We develop $\cal L$ in series on the background, where $f = {\tilde f} = 0$, up to the fourth order


\begin{align}
&
{\cal L} ({\cal F},{\cal G}) = 
{\cal L} ({\cal F}_{\rm B} + \delta {\cal F}, {\cal G}_{\rm B} + \delta {\cal G}) = 
{\cal L} ({\cal F}_{\rm B}, {\cal G}_{\rm B} ) + \left.\frac{1}{n!}\left(\delta {\cal F}\frac{\partial}{\partial {\cal F}} + 
 \delta {\cal G}\frac{\partial}{\partial {\cal G}}\right)^n {\cal L}
\right|_{\rm B} \simeq 
{\cal L} ({\cal F}_{\rm B}, {\cal G}_{\rm B} ) + 
\left. \delta {\cal F}\frac{\partial {\cal L}}{\partial {\cal F}} \right|_{\rm B} +
\left. \delta {\cal G}\frac{\partial {\cal L}}{\partial {\cal G}} \right|_{\rm B} + \nonumber \\
& \left. \frac{\delta {\cal F}^2}{2} \frac{\partial^2{\cal L}}{\partial {\cal F}^2}  \right|_{\rm B} +  
\left. \delta {\cal F}\delta{\cal G} \frac{\partial^2{\cal L}}{\partial {\cal F}\partial {\cal G}}  \right|_{\rm B} +
\left. \frac{\delta {\cal G}^2}{2} \frac{\partial^2{\cal L}}{\partial {\cal G}^2}  \right|_{\rm B} +
\left. \frac{\delta {\cal F}^3}{6} \frac{\partial^3{\cal L}}{\partial {\cal F}^3}  \right|_{\rm B} +
\left. \frac{\delta {\cal F}^2\delta{\cal G}}{2} \frac{\partial^3{\cal L}}{\partial {\cal F}^2\partial{\cal G}}  \right|_{\rm B} + 
\left. \frac{\delta {\cal F}\delta{\cal G}^2}{2}\frac{\partial^3{\cal L}}{\partial {\cal F}\partial{\cal G}^2} \right|_{\rm B} + 
\left. \frac{\delta {\cal G}^3}{6}\frac{\partial^3{\cal L}}{\partial {\cal G}^3}  \right|_{\rm B} + \nonumber \\
&
\left. \frac{\delta {\cal F}^4}{24} \frac{\partial^4{\cal L}}{\partial {\cal F}^4}  \right|_{\rm B} + 
\left. \frac{\delta {\cal F}^3\delta{\cal G}}{6} \frac{\partial^4{\cal L}}{\partial {\cal F}^3\partial{\cal G}}  \right|_{\rm B} +
\left. \frac{\delta {\cal F}^2\delta{\cal G}^2}{3} \frac{\partial^4{\cal L}}{\partial {\cal F}^2\partial{\cal G}^2}  \right|_{\rm B} + \left. \frac{\delta {\cal F}\delta{\cal G}^3}{6} \frac{\partial^4{\cal L}}{\partial {\cal F}\partial{\cal G}^3}  \right|_{\rm B} + 
\left. \frac{\delta {\cal G}^4}{24} \frac{\partial^4{\cal L}}{\partial {\cal G}^4}  \right|_{\rm B}~,
\end{align}

and after some lengthy computation, we explicit the fields up to fourth order

\begin{align}
{\cal L}_4 & = 
- C_1 \left( \frac{1}{2} F_{\rm B} f + \frac{1}{4} f^2 \right) 
- C_2  \left(\frac{1}{2} {\tilde F}_{\rm B} f +\frac{1}{4} f {\tilde f} \right) + 
\frac{D_1}{2}  \left( \frac{1}{2} F_{\rm B} f + \frac{1}{4} f^2 \right)^2  + 
D_2 \left(\frac{1}{2} F_{\rm B} f + \frac{1}{4} f^2 \right)\left( \frac{1}{2} {\tilde F}_{\rm B} f +\frac{1}{4} f {\tilde f} \right) +
\frac{D_3}{2} \left( \frac{1}{2} {\tilde F}_{\rm B} f + \frac{1}{4} f {\tilde f} \right)^2 -
\nonumber \\
&
\frac{M_1}{6} \left(\frac{1}{2} F_{\rm B} f + \frac{1}{4} f^2 \right)^3 - 
\frac{M_3}{2} \left(\frac{1}{2} F_{\rm B} f + \frac{1}{4} f^2 \right)^2 \left( \frac{1}{2} {\tilde F}_{\rm B} f + \frac{1}{4} f {\tilde f} \right) - 
\frac{M_4}{2} \left( \frac{1}{2} F_{\rm B} f + \frac{1}{4} f^2 \right)\left(\frac{1}{2} {\tilde F}_{\rm B} f + \frac{1}{4} f {\tilde f} \right)^2 -
\frac{M_2}{6} \left( \frac{1}{2} {\tilde F}_{\rm B} f + \frac{1}{4} f {\tilde f} \right)^3 +
\nonumber \\
& 
\frac{N_1}{24} \left(\frac{1}{2} F_{\rm B} f + \frac{1}{4} f^2 \right)^4 +
\frac{N_3}{6} \left(\frac{1}{2} F_{\rm B} f + \frac{1}{4} f^2 \right)^3 \left(\frac{1}{2} {\tilde F}_{\rm B} f + \frac{1}{4} f {\tilde f} \right) +
\frac{N_4}{3} \left( \frac{1}{2} F_{\rm B} f + \frac{1}{4} f^2 \right)^2\left(\frac{1}{2} {\tilde F}_{\rm B} f + \frac{1}{4} f {\tilde f} \right)^2 + 
\nonumber \\
&
\frac{N_5}{6} \left(\frac{1}{2} F_{\rm B} f + \frac{1}{4} f^2 \right) \left( \frac{1}{2} {\tilde F}_{\rm B} f + \frac{1}{4} f {\tilde f} \right)^3 + 
\frac{N_2}{24} \left(\frac{1}{2} {\tilde F}_{\rm B} f + \frac{1}{4} f {\tilde f} \right)^4 ~,
\end{align}

having defined

\begin{align}
& 
\left. {\ds \frac{\partial{\cal L}}  {\partial {\cal F}}} \right|_{\rm B}= C_1~, 
\left. {\ds \frac{\partial{\cal L}}  {\partial {\cal G}}}  \right|_{\rm B} = C_2~, 
\left. {\ds \frac{\partial^2{\cal L}}{\partial {\cal F}^2}}  \right|_{\rm B} = D_1~, 
\left. {\ds \frac{\partial^2{\cal L}}{\partial {\cal F}\partial {\cal G}}}  \right|_{\rm B} = D_2~,
\left. {\ds \frac{\partial^2{\cal L}}{\partial {\cal G}^2}}  \right|_{\rm B} = D_3~,
\left. {\ds \frac{\partial^3{\cal L}}{\partial {\cal F}^3}}  \right|_{\rm B} = M_1~,
\left. {\ds \frac{\partial^3{\cal L}}{\partial {\cal F}^2\partial {\cal G}}}  \right|_{\rm B} = M_3~,
\left. {\ds \frac{\partial^3{\cal L}}{\partial {\cal F}\partial {\cal G}^2}}  \right|_{\rm B} = M_4~, \nonumber \\
&
\left. {\ds \frac{\partial^3{\cal L}}{\partial {\cal G}^3}}  \right|_{\rm B} = M_2~,
\left. {\ds \frac{\partial^4{\cal L}}{\partial {\cal F}^4}}  \right|_{\rm B} = N_1~,
\left. {\ds \frac{\partial^4{\cal L}}{\partial {\cal F}^3\partial {\cal G}}}  \right|_{\rm B} = N_3~
\left. {\ds \frac{\partial^4{\cal L}}{\partial {\cal F}^2\partial {\cal G}^2}}  \right|_{\rm B} = N_4~, 
\left. {\ds \frac{\partial^4{\cal L}}{\partial {\cal F}\partial {\cal G}^3}}  \right|_{\rm B} = N_5~,
\left. {\ds \frac{\partial^4{\cal L}}{\partial {\cal G}^4}}  \right|_{\rm B} = N_2~. \nonumber
\end{align}
\end{widetext}

\justifying 
{The linearisation of the Lagrangian shows that interactions are absent at first order; at second, there is interaction between the photon and background field; at third, we know that a photon may split in two photons and two merge into one; at fourth, the photon-photon interaction produces two new photons. }
We return to the index notation and write 

\begin{align}
{\cal L}_4 = &   
- \frac{1}{2} \left(C_1 F_{\rm B}^{\mu\nu} + 
C_2 {\tilde F}_{\rm B}^{\mu\nu}\right) f_{\mu\nu} - 
\frac{1}{4} C_1 f^{\mu\nu}f_{\mu\nu} - \frac{1}{4} C_2 f^{\mu\nu}{\tilde f}_{\mu\nu} +
\nonumber \\
& 
\frac{1}{8} \left ( K_{\rm B}^{\mu\nu\kappa\lambda} +  
2T_{\rm B}^{\mu\nu\kappa\lambda}\right ) f_{\mu\nu} f_{\kappa\lambda} + 
\frac{1}{8} R_{\rm B}^{\mu\nu\kappa\lambda\rho\sigma} f_{\mu\nu} f_{\kappa\lambda} f_{\rho\sigma} + 
\nonumber \\
&
\frac{1}{16} S_{\rm B}^{\mu\nu\kappa\lambda\rho\sigma\omega\tau} f_{\mu\nu} f_{\kappa\lambda} f_{\rho\sigma}f_{\omega\tau}~, 
\label{tevapri}
\end{align}
having posed the following definitions  

\beq
K^{\mu\nu\kappa\lambda}_{\rm B} = 
D_1 F_{\rm B}^{\mu\nu} F_{\rm B}^{\kappa\lambda}
+ D_3 {\tilde F}_{\rm B}^{\mu\nu} {\tilde F}_{\rm B}^{\kappa\lambda}~, 
~~
T^{\mu\nu\kappa\lambda}_{\rm B} = 
 D_2 {\tilde F}_{\rm B}^{\mu\nu}{F}_{\rm B}^{\kappa\lambda} ~, \nonumber
\eeq
where $K^{\mu\nu\kappa\lambda}$ and $T^{\mu\nu\kappa\lambda}_B$ are anti-symmetric in [$\mu\nu$] and [$\kappa\lambda$] but symmetric under the exchange of the pairs [$\mu\nu$] - [$\kappa\lambda$]; 

\begin{align}
R^{\mu\nu\kappa\lambda\rho\sigma}_{\rm B} = & 
D_1 F_{\rm B}^{\mu\nu} \eta^{\kappa\rho} \eta^{\lambda\sigma} \!+\! 
\frac{1}{2} D_3 {\tilde F}_{\rm B}^{\mu\nu} \epsilon^{\kappa\lambda\rho\sigma} \!+\!
\frac{1}{2} D_2 F_{\rm B}^{\mu\nu} \epsilon^{\kappa\lambda\rho\sigma} \!+\!
\nonumber \\
& 
D_2 {\tilde F}_{\rm B}^{\mu\nu} \eta^{\kappa\rho} \eta^{\lambda\sigma} \!-\! 
\frac{1}{6} M_1 F_{\rm B}^{\mu\nu} F_{\rm B}^{\kappa\lambda} F_{\rm B}^{\rho\sigma} \!-\!
\frac{1}{6} M_2 {\tilde F}_{\rm B}^{\mu\nu} {\tilde F}_{\rm B}^{\kappa\lambda} {\tilde F}_{\rm B}^{\rho\sigma} \!-\!
\nonumber \\
&
\frac{1}{2} M_3 F_{\rm B}^{\mu\nu} F_{\rm B}^{\kappa\lambda} {\tilde F}_{\rm B}^{\rho\sigma} \!-\!
\frac{1}{2} M_4 F_{\rm B}^{\mu\nu} {\tilde F}_{\rm B}^{\kappa\lambda} {\tilde F}_{\rm B}^{\rho\sigma}~,
\end{align}

\begin{align}
S^{\mu\nu\kappa\lambda\rho\sigma\omega\tau}_B = &  
\frac{1}{2} D_1 \eta^{\mu\kappa} \eta^{\nu\lambda} \eta^{\rho\omega} \eta^{\sigma\tau} \!+\!
\frac{1}{8} D_3 \epsilon^{\mu\nu\kappa\lambda} \epsilon^{\rho\sigma\omega\tau} \!+\! \nonumber \\
& 
\frac{1}{2} D_2 \eta^{\mu\kappa} \eta^{\nu\lambda} \epsilon^{\rho\sigma\omega\tau} \!-\! 
\frac{1}{2} M_1 F_{\rm B}^{\mu\nu} F_{\rm B}^{\kappa\lambda} \eta^{\rho\omega} \eta^{\sigma\tau} \!-\!  \nonumber \\
&
\frac{1}{4} M_2 {\tilde F}_{\rm B}^{\mu\nu} {\tilde F}_{\rm B}^{\kappa\lambda} \epsilon^{\rho\sigma\omega\tau} \!-\! 
\frac{1}{4} M_3 F_{\rm B}^{\mu\nu} F_{\rm B}^{\kappa\lambda} \epsilon^{\rho\sigma\omega\tau} \!-\! \nonumber \\
&
M_3 F_{\rm B}^{\mu\nu} {\tilde F}_{\rm B}^{\kappa\lambda} \eta^{\rho\omega} \eta^{\sigma\tau} \!-\!  
\frac{1}{2} M_4 F_{\rm B}^{\mu\nu} {\tilde F}_{\rm B}^{\kappa\lambda} \epsilon^{\rho\sigma\omega\tau} \!-\! \nonumber \\
&
\frac{1}{2} M_4 {\tilde F}_{\rm B}^{\rho\sigma} {\tilde F}_{\rm B}^{\omega\tau} \eta^{\mu\kappa} \eta^{\nu\lambda} \!+\! 
\frac{1}{24} N_1 F_{\rm B}^{\mu\nu} F_{\rm B}^{\kappa\lambda} F_{\rm B}^{\rho\sigma} F_{\rm B}^{\omega\tau} \!+\!
\nonumber \\
&
\frac{1}{24} N_2 {\tilde F}_{\rm B}^{\mu\nu} {\tilde F}_{\rm B}^{\kappa\lambda} {\tilde F}_{\rm B}^{\rho\sigma} {\tilde F}_{\rm B}^{\omega\tau} \!+\!
\frac{1}{6} N_3 F_{\rm B}^{\mu\nu} F_{\rm B}^{\kappa\lambda} F_{\rm B}^{\rho\sigma} {\tilde F}_{\rm B}^{\omega\tau} \!+\!
\nonumber \\
& \frac{1}{4} N_4 F_{\rm B}^{\mu\nu} F_{\rm B}^{\kappa\lambda} {\tilde F}_{\rm B}^{\rho\sigma} {\tilde F}_{\rm B}^{\omega\tau} \!+\!
\frac{1}{6} N_5 F_{\rm B}^{\mu\nu} {\tilde F}_{\rm B}^{\kappa\lambda} {\tilde F}_{\rm B}^{\rho\sigma} {\tilde F}_{\rm B}^{\omega\tau}~.
\end{align}

The parameters $C_i,D_i,M_i,N_i$ depend only on the background field. If the latter is constant, the parameters will be constant.
We pose $K^{\mu\nu\kappa\lambda}_{\rm B} +2 T^{\mu\nu\kappa\lambda}_{\rm B} = {\hat Q}^{\mu\nu\kappa\lambda}_{\rm B}$. 

$R^{\mu\nu\kappa\lambda\rho\sigma}_{\rm B}$ is contracted with $f_{\mu\nu}$, $f_{\kappa\lambda}$ and $f_{\rho\sigma}$; the sole surviving
 components is anti-symmetric  in [$\mu\nu$], [$\kappa\lambda$], and [$\rho\sigma$], 
 and named ${\tilde R}^{\mu\nu\kappa\lambda\rho\sigma}_{\rm B}$. Similarly, it occurs for ${\tilde 
 S}^{\mu\nu\kappa\lambda\rho\sigma\omega\tau}_{\rm B}$.  
 
The background-dependent tensorial coefficients, $R_{\rm B}$ and $S_{\rm B}$, in the 
3- and 4-photon vertices of the Lagrangian density, Eq. (\ref{tevapri}), respectively, present negative mass dimensions: $R_{\rm B}$ with dimension (-2) and  $S_{\rm B}$, with dimension (-4) - imply that the model, in its quantum-field theoretical version, is non-renormalisable. This is not an issue as we deal with a classical effective field theory that holds below a well defined cut-off, fixed by the physics we investigate. In this case, the effective field theory stretches at most up to the GeV scale, since it is a purely photonic model, without available energy to excite the weak boson mediators. 

The field equations are obtained by the variation of the Lagrangian with respect to the 4-potential $a_\mu$ of the photon field 
$ f_{\mu\nu} = \partial_\mu a_\nu - \partial_\nu a_\mu$.

We observe that $R$ and $S$ are contracted with $f$ 

\begin{align}
{\cal L}_4= & 
- \frac{1}{2} \left(C_1 F_{\rm B}^{\mu\nu} + C_2 {\tilde F}_{\rm B}^{\mu\nu}\right) f_{\mu\nu} 
- \frac{1}{4} C_1 f^{\mu\nu}f_{\mu\nu} -\frac{1}{4} C_2 f^{\mu\nu}{\tilde f}_{\mu\nu} \nonumber \\ 
& + \frac{1}{8}{\hat Q}^{\mu\nu\kappa\lambda}_B f_{\mu\nu} f_{\kappa\lambda} + 
\frac{1}{8} R_{\rm B}^{\mu\nu\kappa\lambda\rho\sigma} f_{\mu\nu} f_{\kappa\lambda} f_{\rho\sigma} +
\nonumber \\
&
\frac{1}{16} S_{\rm B}^{\mu\nu\kappa\lambda\rho\sigma\omega\tau} f_{\mu\nu} f_{\kappa\lambda} f_{\rho\sigma}f_{\omega\tau}~, 
\label{lagrangian}
\end{align}

and thereby, we have

\begin{align}
{\cal L}_4= &  
- \left(C_1 F_{\rm B}^{\mu\nu}+ C_2 {\tilde F}_{\rm B}^{\mu\nu}\right)  \partial_\mu \delta a_\nu 
-  C_1 f^{\mu\nu}  \partial_\mu \delta a_\nu 
-  C_2  {\tilde f}^{\mu\nu} \partial_\mu \delta a_\nu +  \nonumber \\  
& \frac{1}{2}{\hat Q}^{\mu\nu\kappa\lambda}_B  f_{\kappa\lambda}\partial_\mu \delta a_\nu + 
 R_{\rm B}^{\mu\nu\kappa\lambda\rho\sigma}  f_{\kappa\lambda} f_{\rho\sigma}\partial_\mu \delta a_\nu +
\nonumber \\
&
 S_{\rm B}^{\mu\nu\kappa\lambda\rho\sigma\omega\tau} f_{\kappa\lambda} f_{\rho\sigma}f_{\omega\tau}\partial_\mu \delta a_\nu~. 
\label{work1}
\end{align}

Integrating by parts, we get the field equations

\begin{align}
&  \partial_\mu \left (C_1 f^{\mu\nu} + C_2  {\tilde f}^{\mu\nu} \right) \delta a_\nu - \frac12\partial_\mu \left( {\hat Q}^{\mu\nu\kappa\lambda}_B  f_{\kappa\lambda} \right ) \delta a_\nu \nonumber \\
 - &\partial_\mu \left ( R_{\rm B}^{\mu\nu\kappa\lambda\rho\sigma}  f_{\kappa\lambda} f_{\rho\sigma}\right) \delta a_\nu 
- \partial_\mu  \left ( S_{\rm B}^{\mu\nu\kappa\lambda\rho\sigma\omega\tau} f_{\kappa\lambda} f_{\rho\sigma}f_{\omega\tau} \right) \delta a_\nu =  \nonumber \\ 
& \partial_\mu \left(C_1 F_{\rm B}^{\mu\nu}+ C_2 {\tilde F}_{\rm B}^{\mu\nu}\right)  \delta a_\nu ~, 
\label{feq1}
\end{align}
that is 

\begin{align}
&  \partial_\mu \left (C_1 f^{\mu\nu} + C_2  {\tilde f}^{\mu\nu} \right)- \frac12\partial_\mu \left( {\hat Q}^{\mu\nu\kappa\lambda}_B  f_{\kappa\lambda} \right )- \partial_\mu \left ( R_{\rm B}^{\mu\nu\kappa\lambda\rho\sigma}  f_{\kappa\lambda} f_{\rho\sigma}\right)&  \nonumber \\
- &\partial_\mu  \left ( S_{\rm B}^{\mu\nu\kappa\lambda\rho\sigma\omega\tau} f_{\kappa\lambda} f_{\rho\sigma}f_{\omega\tau} \right)  = \partial_\mu \left(C_1 F_{\rm B}^{\mu\nu}+ C_2 {\tilde F}_{\rm B}^{\mu\nu}\right) ~.  
\label{feq2}
\end{align}

Adding an external current and multiplying each term by $f_{\nu\alpha}$, we obtain

\begin{align}
&  \partial_\mu \left (C_1 f^{\mu\nu} + C_2  {\tilde f}^{\mu\nu} \right)f_{\nu\alpha}- \frac12\partial_\mu \left( {\hat Q}^{\mu\nu\kappa\lambda}_B  f_{\kappa\lambda} \right )f_{\nu\alpha}- &  \nonumber \\
&\partial_\mu \left ( R_{\rm B}^{\mu\nu\kappa\lambda\rho\sigma}  f_{\kappa\lambda} f_{\rho\sigma}\right)f_{\nu\alpha}-\partial_\mu  \left ( S_{\rm B}^{\mu\nu\kappa\lambda\rho\sigma\omega\tau} f_{\kappa\lambda} f_{\rho\sigma}f_{\omega\tau} \right)f_{\nu\alpha}  =\nonumber \\ 
&   
\partial_\mu \left(C_1 F_{\rm B}^{\mu\nu}+ C_2 {\tilde F}_{\rm B}^{\mu\nu}\right) f_{\nu\alpha} + j^\nu f_{\nu\alpha}~. 
\label{feq3}
\end{align}
  
After considerable manipulation, we have the density of the photon energy-momentum tensor (EMT), which for zero external current, is given by 

{\begin{align}
    {\theta^\mu}_\alpha= &C_1 f^{\mu\nu}f_{\nu\alpha} - \frac12 {\hat Q}^{\mu\nu\kappa\lambda}_B  f_{\kappa\lambda} f_{\nu\alpha}- \frac34 R_{\rm B}^{\mu\nu\kappa\lambda\rho\sigma}  f_{\kappa\lambda} f_{\rho\sigma}f_{\nu\alpha}&  \nonumber \\
- &S_{\rm B}^{\mu\nu\kappa\lambda\rho\sigma\omega\tau} f_{\kappa\lambda} f_{\rho\sigma}f_{\omega\tau}f_{\nu\alpha}  +{\delta^\mu}_\alpha \left(\frac{1}{4} C_1 f^{\beta\nu}f_{\beta\nu} - \frac{1}{8}{\hat Q}^{\beta\nu\kappa\lambda}_B f_{\beta\nu} f_{\kappa\lambda}\right.\nonumber \\ 
 -
&\frac{1}{8} R_{\rm B}^{\beta\nu\kappa\lambda\rho\sigma} f_{\beta\nu} f_{\kappa\lambda} f_{\rho\sigma} -
\left.\frac{1}{16} S_{\rm B}^{\beta\nu\kappa\lambda\rho\sigma\omega\tau} f_{\mu\nu} f_{\kappa\lambda} f_{\rho\sigma}f_{\omega\tau}\right)~.
\label{right2}
\end{align}}

We recall the physical meaning of the photon energy-momentum tensor density [Jm$^{-3}$], in SI units:  
$\theta^{0}_{\ 0}$ = energy density, $\theta^{0}_{\ k}$ = energy flux divided by $c$ along the $k$ direction, 
$\theta^{k}_{\ 0}$ =  momentum density through the orthogonal surface to $k$, multiplied by c. 
The density of the photon EMT variation in SI units [Jm$^{-4}$] and reinserting $\mu_0$
is given by 

{\begin{align}
& \partial_{ \alpha} {\theta^ \alpha}_\tau = 
\underbrace{
- \frac{1}{\mu_0}\partial_\alpha C_1 {F}^{\alpha\nu} f_{\nu\tau}
}_{\text{Maxwellian term}} 
\underbrace{-\frac{1}{\mu_0}\partial_\alpha C_2\Tilde{F}^{\alpha\nu}f_{\nu\tau} }_{\text{non-Maxwellian linear term}} \nonumber \\ 
& \underbrace{
+ \frac{1}{4\mu_0} \left( \partial_\tau C_1 \right) f^{ \alpha\nu}f_{ \alpha\nu} + \frac{1}{4\mu_0} \left( \partial_\tau C_2 \right) \Tilde{f}^{ \alpha\nu}f_{ \alpha\nu} - \frac{1}{8\mu_0^2} \left( \partial_\tau {\hat Q}^{ \alpha\nu\kappa\lambda} \right )f_{ \alpha\nu}f_{\kappa\lambda} }_{\text{second order non-linear terms}} 
 \nonumber\\
& \underbrace{ - \frac{1}{8\mu_0^3}\left(\partial_\tau R^{ \alpha\nu\kappa\lambda\rho\sigma}\right) f_{ \alpha\nu} f_{\kappa\lambda} f_{\rho\sigma} - \frac{1}{16\mu_0^4}\left(\partial_\tau S^{ \alpha\nu\kappa\lambda\rho\sigma\omega\xi} \right)f_{ \alpha\nu}f_{\kappa\lambda} f_{\rho\sigma}f_{\omega\xi} }_{\text{third and fourth non-linear terms}}~.
\label{right4}
\end{align}}

The energy density for the photon field can be obtained by taking the (0,0) component of the EMT, named $\cal E$, which at the second order, in SI units, gives

\begin{align}
    {\cal E} = \frac{C_1}{2\mu_0}\left(\frac{\vec{e}^{~2}}{c^2}+\vec{b}^2\right)+\frac{D_1}{2{\mu_0}^2}\left[\left(\frac{\vec{E}\cdot\vec{e}}{c^2}\right)^2-\left(\vec{B}\cdot\vec{b}\right)^2\right]~.
\end{align}

The energy density time variation is 
{\begin{align}
    \frac{\partial\mathcal{E}}{\partial t}=&\frac{1}{2\mu_0}\left(\frac{\vec{e}^{~2}}{c^2}+\vec{b}^2\right)\partial_t C_1+\frac{1}{2{\mu_0}^2}\left[\left(\frac{\vec{E}\cdot\vec{e}}{c^2}\right)^2-\left(\vec{B}\cdot\vec{b}\right)^2\right]\partial_t D_1 &\nonumber\\
    &+ \frac{D_1}{2{\mu_0}^2} \left(\frac{\Vec{e}^2}{c^4}\partial_t  \Vec{E}^2-\Vec{b}^2\partial_t \Vec{B}^2\right)~.
\end{align}}


Due to the experimental correspondence wave-single photon \cite{aspect-grangier-1987}, these two quantities represent also respectively the frequency and the frequency shift. We now take into account two specific scenarii, namely the Euler-Heisenberg and the Born-Infeld cases. 

\begin{widetext}

\section {The case of Euler-Heisenberg and Born-Infeld theories}

The non-conservation of photon energy already appears at second order for which the EH and BI Lagrangians, represented by our generalised formalism, look as \cite{Sorokin-2022}

\begin{align}
&{\cal L}_{\rm EH} = - \frac{1}{4\mu_0}{F}_{\sigma\tau}{F}^{\sigma\tau} + \beta
\left[\left({F}_{\sigma\tau}{F}^{\sigma\tau}\right)^2 + 7 \left({G}_{\sigma\tau}{F}^{\sigma\tau}\right)^2\right]~,
\label{laeh}\\
&{\cal L}_{\rm BI} = - \frac{1}{4\mu_0}{F}_{\sigma\tau}{F}^{\sigma\tau} + \frac{1}{32b} 
\left[\left({F}_{\sigma\tau}{F}^{\sigma\tau}\right)^2 + \left({G}_{\sigma\tau}{F}^{\sigma\tau}\right)^2\right]~,
\label{labi}
\end{align}
where $b$ is a scale parameter and $\beta$ is a constant depending on the electron mass and fine structure constant 
$ \beta = {2\alpha^2}/{45{m_e}^4}$. 
 
For the Euler-Heisenberg case, we get the coefficients 
$C_1=1+2\beta\mathcal{F}_B, C_2=14\beta\mathcal{G}_B, D_1=2\beta, D_3=14\beta, D_2=0$,   
and the energy density is 

\beq
    {\cal E}_{\rm EH} = \frac{1}{2\mu_0}\left[1+\frac{\beta}{\mu_0}\left(\frac{\vec{E}^{~2}}{c^2}-\vec{B}^2\right)\right]\left(\frac{\vec{e}^{~2}}{c^2}+\vec{b}^2\right) +
   \frac{\beta}{{\mu_0}^2}\left(\left[\frac{\vec{E}\cdot\vec{e}}{c^2}\right)^2-\left(\vec{B}\cdot\vec{b}\right)^2\right]~.
\label{EEH}
\eeq

We repeat the same procedure for the Born-Infeld model. Starting by its coefficients
$ C_1=1+ 1/ 16b\mathcal{F}_B, C_2=1/16b\mathcal{G}_B, D_1=D_3= 1/16b, D_2=0$, 
the energy density is given by
\beq
        {\cal E}_{\rm BI} = \frac{1}{2\mu_0}\left[1+\frac{1}{16\mu_0 b}\left(\frac{\vec{E}^{~2}}{c^2}-\vec{B}^2\right)\right]\left(\frac{\vec{e}^{~2}}{c^2}+\vec{b}^2\right) +
    \frac{1}{{16\mu_0}^2b}\left[\left(\frac{\vec{E}\cdot\vec{e}}{c^2}\right)^2-\left(\vec{B}\cdot\vec{b}\right)^2\right]~.
\eeq

Now, we compare the two energy density variations

{\begin{align}
    \frac{\partial\mathcal{E}_{EH}}{\partial t}= \frac{\beta}{2\mu_0^2}\left(\frac{\vec{b}^2}{c^2}\partial_t\Vec{E}^2-\frac{\vec{e}^{~2}}{c^2}\partial_t \Vec{B}^2\right) 
    + \frac{3\beta}{2{\mu_0}^2} \left(\frac{\Vec{e}^2}{c^4}\partial_t  \Vec{E}^2-\Vec{b}^2\partial_t \Vec{B}^2\right)~,
\end{align}

\begin{align}
      \frac{\partial\mathcal{E}_{BI}}{\partial t}=&\frac{1}{64b\mu_0^2}\left(\frac{\vec{b}^2}{c^2}\partial_t\Vec{E}^2-\frac{\vec{e}^{~2}}{c^2}\partial_t \Vec{B}^2\right) 
    + \frac{3}{64b{\mu_0}^2} \left(\frac{\Vec{e}^2}{c^4}\partial_t  \Vec{E}^2-\Vec{b}^2\partial_t \Vec{B}^2\right)~.
\end{align}}

\end{widetext}

Differentiating BI from EH frequency shifts is not an easy undertaking. Furthermore, in a laboratory setting when using an interferometric cavity, other components than $\theta^0_{~0}$ have to computed. The discrimination of frequency changes from velocity changes, cavity wall displacements and accumulated phase changes, require that the entire EMT is to be computed.   

\section{Discussions and perspectives}

\subsection{Discussion}

We call into question the issue of the Poincar\'e Symmetry Violation (PSV).The fundamental symmetries of space-time (isotropy and homogeneity) can be expressed in the language of group theory. These symmetries are demonstrated by the generators of the Poincar\'e group which adds translation transformations to the Lorentz sub-group of transformations, namely rotations in space and boosts. A Poincar\'e symmetry Violation implies that the full Poincar\'e group breaks down. 

In the case of a space-time constant electro-magnetic background, the continuity equation, Eq. (\ref{right4}), shows that the energy-momentum tensor is conserved. This is a consequence, via Noether's theorem, of the Lagrangian density, Eq. (\ref{work1})
not exhibiting an explicit space-time dependence. The Lorentz sub-group of the Poincar\'e symmetry is violated in
the active sense. Active (particle) transformations are transformations of the physical fields, whereas passive (observer) transformations are merely relabellings of points. Active transformations act on the physical system under consideration without touching the background. Conversely,  the passive viewpoint applies when both the physical system and the background are transformed. Active and passive transformations coincide when the metric components rest the same, but differ in presence of a tensorial anisotropic field \cite{Duffy-2022}. If fundamental theories must have the same symmetries as space-time itself, differing between active and passive becomes unnecessary. In the case of a space-time varying background fields yield non-vanishing components of the Q-, R- and S-tensors present in Eq. (\ref{work1}). So, symmetry under space-
time rotations is broken, which means Lorentz symmetry
violation in the active sense. This can be verified by explicitly computing the canonical angular momentum current tensor density of the photon in this system through Noether's theorem

\begin{align}
    \mathcal{M}^{\alpha\mu\nu}=& i\left(x^\mu T^{\alpha\nu} - x^\nu T^{\alpha\mu}\right)- \mathcal{S}^{\alpha\mu\nu}~,
   \end{align}
where $T$ is the canonical energy-momentum tensor, and 

{\begin{align}
   & \mathcal{S}^{\alpha\mu\nu}=\mathcal{J}^{\alpha\mu}a^\nu -\mathcal{J}^{\alpha\nu}a^\mu + C_1 (f^{\alpha\mu}a^{\nu}- f^{\alpha\nu}a^{\mu}) + \nonumber\\
    & C_2 (\tilde{f}^{\alpha\mu}a^{\nu}- \tilde{f}^{\alpha\nu}a^{\mu}) + \frac12\left( Q^{\alpha\nu\rho\sigma}f_{\rho\sigma}a^\mu - Q^{\alpha\mu\rho\sigma}f_{\rho\sigma}a^\nu \right)
\end{align}}
is the spin density tensor, being $\mathcal{J}^{\alpha\mu}=\left(C_1 F_{\rm B}^{\alpha\mu} + C_2 {\tilde F}_{\rm B}^{\alpha\mu}\right)$. 

The presence of vector and tensor anti-symmetric quantities in the background - space-time variable or not - implies that the energy-momentum tensor is no longer symmetric, Eq. (\ref{right2}). In this situation, the Poynting vector, $\theta^{~i0}$, and the momentum density carried by the propagating field, $\theta^{~0i}$, are different. We do not report here, but it can be shown that, from the time variation of the momentum density $\theta^{~0i}$ - which indicates the breaking of space translational symmetry - the effects of the (variable) background contribute asymmetrically to the usual Lorentz force density.

The fact that we are dealing with a Poincar\'e symmetry violating model brings into question multiple issues, but one in particular: the Casimir operators\footnote{The two Casimir operators are $P^2=P_\mu P^\mu$ and $W^2=W_\mu W^\mu$, where $P$ is momentum, and $W$ is the  Pauli-Lubanski pseudo-vector. In the rest frame, we get $P^2=mc^2$, that is mass for the first Casimir operator. Instead $W^2=P_\mu P^\mu$.
has to do with spin for massive particles and helicity for massless particles.} of the group. Indeed, the Poincar\'e group possesses two invariant operators, which are associated to the mass and the helicity/spin of the states that carry its unitary irreducible representations. In the case described above, it is safe to assume that at least one of these quantities is no longer an invariant, thereby putting into question the formal definition of mass and/or spin. Could a Poincar\'e symmetry violation lead to emerging non-constant masses? What would a variable spin imply physically? How would this fundamentally affect the structure of space-time? 

\subsection{Perspectives}

Herein, we remind the discussion we have presented in \cite{helayelneto-spallicci-2019,spallicci-etal-2021,spallicci-sarracino-capozziello-2022,sarracino-spallicci-capozziello-2022}, where the reader can find the different cosmological models and the detailed calculations for massive and the SME formalisms. 

The sizing of the frequency shift is not straightforward, as many factors enter into play. 
First, it depends on the amount of distance crossed by the photon. 
Second, the photon crosses several EM fields for the source to the observer.  
The magnetic field in the Milky Way has a strength of around 0.5 nT and it has regular and fluctuating components of comparable strengths. In the external spiral galaxies the fields resemble presumably that in our own Galaxy.  In elliptical galaxies, it is supposed that only the fluctuating components survive. For the Inter-Galactic Medium, reliable conclusion cannot be drawn neither, but supposing a strength of nT order of magnitude, it is the safest hypothesis we can make \cite{Ferriere-2015,Han-2017} 
The inter-stellar and inter-galactic media are good electric conductors, such that magnetic fields are frozen in the plasma. Thereby, the electric field is given by
${\vec E} \propto {\vec v}_p \times {\vec B}$,
where $v_p$ is the plasma velocity.
In general, $v_p \ll c$, thus $E \ll B$ and thereby potentially negligible. This assumption may not hold locally though, and photons may pass through intense electric fields.
Third, the orientation of the (inter-)galactic fields with respect to the photon propagation vector. 
Fourth, the effect and the alignment of the EM fields encountered by the photon. An EM field may determine a shift towards the blue, while another towards the red. 
Fifth, we the evaluation value of the EM field of a photon is a fundamental issue.  

For the SME, we had some preliminary assessments \cite{helayelneto-spallicci-2019}. 
The static shift varies in the range 0.01 - 10\% of the expansion shift. Applied to SNeIa, the static shift evidently differs from one SNIa to the other, as the above mentioned variables are many. But this is a great advantage as it allows a great freedom to apply the model to any SNIa in the luminosity-red shift plot.  

But we can reverse our approach and ask which frequency shift is necessary to explain specific phenomena. 
In \cite{spallicci-etal-2021,spallicci-sarracino-capozziello-2022, sarracino-spallicci-capozziello-2022}, we found agreement between SNe luminosity and red shift distances by recasting the observed red shift as a combination of the expansion shift and a static shift due to a massive theory (physical mass as in dBP or effective as in the SME). Thereby, we could get rid of the necessity of the accelerated expansion. Acceleration fits in the Friedmann-Lema\^itre-Robertson-Walker cosmology, which is not more than an idealised solution of general relativity, having  disregarded other interactions. Dark energy is not an observational result, but it is a plausible interpretation, among others. The real result of the observations is the detection of non-linear dependence of the registered energy current density of SNe Ia with respect to red shift. 

Let $z_{\rm C}$ be the red-shift due to expansion of the universe and $z_{\rm S}$ be the shift due to GNLEM. 
We have then $
1 + z = (1+z_{\rm S}) (1+z_{\rm C}) \simeq 1 + z_{\rm S} + 1 + z_{\rm C}  + z_{\rm S} z_{\rm C}$.

Supposing that the $z_{\rm S}$ shifts, additional to the red-shift due to expansion, be negligible for astrophysics at large scale, they would remain of relevance for the foundations of EM.

Another thread concerns the testing of these effects. The averaged value of the expansion is 70 km/s per Mpc, which corresponds to $2.3 \times 10^{-18}$ m/s per metre \cite{capozziello-benetti-spallicci-2020,spallicci-benetti-capozziello-2022}. Translated as static frequency shift, it becomes  $7.7 \times 10^{-27} \Delta f/f$ per metre. This is to be considered as upper limit, as we intend to add an effect to the expansion, and not to deny the latter. But the state of art of interferometry, for which the measurement of these shifts is absolutely a challenge. 

But leaving aside, these additional shifts, we can ask ourselves another question. Can we test expansion? The formidable technical challenge of measuring the Hubble-(Humason-)Lema\^itre constant of $2.3 \times 10^{-18}$ m/s per metre is not the only hurdle. We have to face the prediction that expansion does not occur at small (below galactic) scale. Therefore, a non-null result in terrestrial or spatial laboratories would be of great significance, as it would prove that frequency shifts of static nature do exist. 

The static frequency shift has recently received attention for the James Webb Space Telescope data  \cite{Gupta-2023}. We emphasise that our GNLEM shift, as the massive and SME theories or even the Maxwellian theory shifts, are computed out of established theories and not out of {\it ad hoc} assumptions, as those reviewed in \cite{LopezCorredoira-Marmet-2022,Marmet-2018}. \\

A further perspective is offered by the reinterpretation of the rotation curves of stars and gasses in galaxies through frequency shifts. If the Doppler shift generated by baryonic matter rotation were accompanied by a shift due to the magnetic  field, acting as background, recurring to dark matter to explain the non-Keplerianity of the curves would be undermined. 

These frequency shifts, if applicable to a quantum system, should semi-classically lead to small divergences from the expected spectral emissions of atoms \cite{Carley-Kiessling-2006, Carley-Kiessling-Perlick-2019}.  

Spin helicity could be not conserved in curved space-time even when dealing with Maxwellian electro-magnetism \cite{Galaverni-Gionti-2021}. In the frame of GNLEM and generally speaking of Extended Theories of Electro-Magnetism, when exhibiting Poincar\'e symmetry violations, helicity might be affected too. These violations are also sought in $\beta$ radiative processes \cite{Gronau-Grossmann-Pirjol-Ryd-2002}. 

Another avenue is constituted by phenomenology in condensed matter systems and especially supraconductivity. Condensed maatter models deal with translational symmetry breaking as well as emerging masses. In fact, they are highly propice systems to study Lorentz symmetry violations and topological theories \`a la Chern-Simons. It could be of interest to analyse whether a split between photons and background were to produce new insights. 

\section{Results}
 
We have shown that GNLEM theories predict a static frequency shift due to the non-conservation of the energy-momentum tensor, occurring when the background is space-time dependent, that we have computed for the Euler-Heisenberg and Born-Infeld theories.
The GNLEM shift, like the other static shifts coming from massive photon and SME frameworks, can be added to the expansion shift providing options for reading astrophysical data or for reading laboratory measurements.
We have argued on the asymmetry of the energy-momentum tensor and on the possible emergence of a mass due to the Poincar\'e symmetry breaking \cite{Gupta-Jaeckel-Spannowsky-2023}, discussing the nature of the latter for non-linear theories and the physical implications of said violations.

\section*{Acknowledgements}

ADAMS is grateful to UERJ (M. A. Lop\'es Capri, V. E. Lemos Rodino hosts for 2019-2020) and the French Consulate in Rio de Janeiro for his nomination as 'Chaire Fran\c{c}aise'. Part of this work was carried out at the D. Poisson in Tours Inst. (M. Chernodub, host) in 2021-2022 during a CNRS funded half-sabbatical. All authors thank the comments by L. Ospedal Prestas Rosas.  

\bibliographystyle{apsrev} 

\end{document}